# THE SUB- AND QUASI-CENTURIAL CYCLES IN SOLAR AND GEOMAGNETIC ACTIVITY DATA SERIES

Boris Komitov[1], Stefano Sello[2], Peter Duchlev[1], Momchil Dechev[1], Kaloyan Penev[3] and Kostadinka Koleva[1]

[1] Institute of Astronomy, 72 Tsarigradsko Chaussee, Sofia 1784, Bulgaria
[2] Mathematical and Physical Models, Enel Research, Via Andrea Pisano 120, 56122 Pisa - Italy;
[3] Harvard University, Mail Stop 10, 60 Garden Street, Cambridge, MA 02138

ABSTRACT

The subject of this paper is the existence and stability of solar cycles with durations in the range of 20–250 years. Five types of data series are used: 1) The Zurich series (1749-2009 AD), the mean annual International sunspot number *Ri*, 2) The Group sunspot number series *Rh* (1610-1995 AD), 3) The simulated extended sunspot *Rsi* number from Extended time series of Solar Activity Indices (ESAI) (1090-2002 AD), 4) The simulated extended geomagnetic *aa*-index from ESAI (1099-2002 AD), 5) The Meudon filament series (1919-1991 AD) (it is used only particularly). Data series are smoothed over 11 years and supercenturial trends are removed. Two principally independent methods of time series analysis are used: the T-R periodogram analysis (both in the standard and "scanning window" regimes) and the wavelet-analysis. The obtained results are very similar. It is found that in all series a strong cycle with mean duration of 55-60 years exists. It is very well expressed in the 18th and the 19th centuries. It is less pronounced during the end of the 19th and the beginning of the 20th centuries. On the other hand a strong and stable quasi 110-120 years and ~200-year cycles are obtained in all of these series except in *Ri*. In the last series a strong mean oscillation of ~ 95 years is found, which is absent in the other data sets. The analysis of the ESAI (1090-2002 AD) proved that the quasi century cycle has a relatively stable doublet (~80 and ~120 years) or triplet (~55-60, 80 and 120 years) structure during the last ~900 years. An interesting feature in all series is the existence of significant ~29-year cycle after the last centurial Gleissberg-Gnevishev's minimum (1898-1923 AD). Most probably the different types of oscillations in the sub-century and century period range correspond to cycles of different classes of active regions. The high importance of the long term solar activity dynamics for the aims of solar dynamo modelling and predictions was emphasized. Some of the Sun-Earth relationship aspects of these results are briefly discussed.

*1. Introduction*

The "conventional" point of view is that the solar flare events generally follow the overall sunspot activity. The typical picture is that the flare maximum in Schwabe-Wolf's cycle follows 1.5-2 years after the main sunspot maximum.

Moreover, the correlation between the international sunspot index *Ri* and solar flare activity is far less convincing than for many others solar phenomena. As it is shown in Table 1 (Section 4), the relationship between the overall sunspot and flare activity is clearly fading from the relatively weaker flare class C to the strongest class X (soft-X-ray flares classification, Crosby(2008)). The very high coefficient of correlation of +0.975 between *Ri* and the radio-index F10.7 indicates that according the Snedekor-Fisher's *F*-test more than 95% of the F10.7 changes follow this relationship during the last 30 years. On other hand, the coefficient of correlation between *Ri* and the strongest flare class X is + 0.449. This implies that only 25% of the X class flares could be related to the overall sunspot activity. That is an indication that the most powerful solar flares are caused by specific physical processes in the solar atmosphere and their dynamics is very independent from the overall sunspot activity.

This conclusion agrees with the results in many recent studies. For solar cycle No 23 Kiliic (2009) found that no significant short-term periodicities (in the range close to a month) in the flare activity exist. The author pointed out a few well-expressed cycles with duration between 23.1 and 36.4 days for the *Ri* data during the studied period. In a new study (Komitov et al, 2010- a study in progress) it is pointed out that the relative and absolute number of powerful solar flares of class X is significantly higher in the sunspot cycle No 20 than in the next two ones No 21 and 22, which are much stronger in the sunspot activity. It is also pointed out that the relative part of the powerful flare classes M and X is greater in the late phases of the sunspot cycles. Such solar cycle phase asymmetry has also been found for the related to the solar flares coronal mass ejections (CME) and Forbush decreases (Kane, 2008a). The sharp increase and extraordinarily high levels of flare activity during the so called "Halloween storms" (October-November, 2003) occurred during average levels of sunspot activity without any strong corresponding increase of $Ri$ (Lilensten and Bornarel, 2006). Series of strong M- and X–class flares have occurred also in lower sunspot level conditions during 2004-2006 AD.

These facts strongly stress the specific dynamics of the strong solar flare classes in comparison to the relative short time scales, of the order of the 11-year Schwabe –Wolf's cycle or less. So, the assumption that in longer time scales (decadal, century or super-century) there could be significant differences between the sunspot $Ri$ index and strong solar M and X class flare dynamics seems very reliable. The problem of the long-term solar flare dynamics and their connection to other solar activity is very important. The understanding of that could help us better understand the physical conditions and generation of the strong flares and "mega-flares" like the "Carrington storm" from 1859 AD (Cliver, 2006) or the "Halloween storms" (October-November, 2003). A successful detection of statistically important long-term cycles, which are closely related to the flare activity could also be a starting point for building time series models and predictions for possible risk epochs of solar flares enhancement. The next very important question is: are there some long-term cycles and how stable are they?

Systematic observations of solar soft-X ray flares are available since August 1975 as a part of the GOES satellites programs. There are additional earlier observations from March 1968, up to February 1974 by the SOLRAD satellite. Regular optical H-alpha flare observations are available since 1936 AD (Link and Kleczek, 1949; Kleczek, 1952a). However according Kleczek (1952b) a serious lack of data during the World War 2 is possible in these data. There are homogenous data since 1976 AD published by the National Geophysical Data Center. A significant part of that data is based on the observations in the Skalnaté Pleso and Kandilli observatories and their analysis (Knoška and Petrásek, 1984; Ataç and Özgüç, 1998). Time series of the coronal mass ejections (CME) events connected to the solar flares are even shorter. Since the 1980s there are two relatively longer periods of regular CME observations: 1984-1989 (SMM satellite) and since January 1996 (SOHO satellite).

Thus there are not enough long instrumental series of direct flare observations to clearly establish long-term flare activity cycles with durations of 30-40 years or more. For the long-term tendencies, as a first step, one could use sunspot, coronal or geomagnetic annual data series such as the Zurich series, the Groups sunspot number (1610 -1995 AD) (Hoyt and Schatten, 1998), as well as some historical or "synthetic" data like the Catalog of the Middle Latitude Aurora (MLA) events (1000- 1900 AD) (Křivský and Pejml, 1988), the simulated sunspot $Rsi$ number and geomagnetic $aa$-index from the Extended time series of Solar Activity Indices (ESAI) (Nagovitsyn, 1997, 2006; Nagovitsyn et al. 2004).

It is a fact that the most powerful solar flares could generate high energy particles ($E \geq 100$ MeV). A part of the generated in the stratosphere "cosmogenic" isotopes atoms ($^{10}$Be, $^{14}$C etc.) are of solar flare origin in addition to the main part, generated by the galactic cosmic rays (Usoskin et al., 2006). This is why the study of the "cosmogenic" isotopes time series is also interesting as a next step.

Strong quasi- 53-55, 60-67 and 115-120 years cycles have been found in the "Greenland $^{10}$Be data series during the epoch 1423-1985 AD (Komitov et al., 2003). These results have been confirmed recently by Komitov and Kaftan (2004). A strong quasi 62-year cycle has been detected also in the time series of MLA-events from the catalog of Křivský and Pejml (1988) during the 18$^{th}$ and 19$^{th}$ century. The extremes are in very good agreement with the corresponding extrema of the cycle with the same duration in the "Greeenland" $^{10}$Be data (Komitov, 2009). The fact that the MLA events are connected to the flare activity of the Sun, including CMEs , suggests that a ~ 60-year cycle should exist in the solar flares dynamics, too. Weak, but statistically significant traces of 50-70 years cycles have been detected in Zurich ($Ri$) and Group sunspot number (GSN) series (Komitov and Kaftan, 2003). A strong "doublet" corresponding to cycles with durations of 114 and 144 years has been found in the synthetic north-south sunspot area asymmetry (Komitov, 2010). A similar result on the basis of the analysis of Greenwich and Solar Optical Observing Network sunspot group data has been found by Javaraiah et al. (2005), who determined its duration to be 113±5 years. A cyclic type trend, with a period of ~117 years has been also found in the Meudon solar filament series (1919-1989 AD) (Duchlev, 2001).

All these results indicate that there are solar activity processes for which the cyclic variations of ~50-60 and ~120 years are important features. They are related to the flares and corresponding active centers and sunspot groups rather than to the overall sunspot activity and the *Ri*-index.

The main aim of this study is a more detailed analysis of the problem of existence and stability of cycles with duration in the range of 20 – 250 years. The analysis uses different solar and geomagnetic data sets of instrumental and "historical" type. Two principally independent methods for time series analysis are used: a) the T-R periodogram analysis (standard (Komitov, 1986, 1997) and "moving window" regime (Bonev *et al.*,2004)); b) wavelet analysis (Torrence and Compo, 1998; Ranucci and Sello, 2004).

*2. Data and Methods*

For the purposes of this study, five sets of direct and indirect data series with annual time step resolution are used:
1. The mean annual Group sunspots number (*GSN*) (1610-1995 AD) (*ftp://ftp.ngdc.noaa.gov/STP/SOLAR_DATA/SUNSPOT_NUMBERS/GROUP_SUNSPOT_NUMBERS/*).
2. The international sunspot number *Ri* index series (1749-2009 AD) (*ftp://ftp.ngdc.noaa.gov/STP/SOLAR_DATA/SUNSPOT_NUMBERS/INTERNATIONAL/*)
3. The simulated international sunspot (*Rsi*) number from ESAI (1090–2002 AD) (Nagovitsyn, 1997, Nagovitsyn et al., 2004) (*http://www.gao.spb.ru/database/esai*)
4. The simulated geomagnetic *aa*-index from the ESAI (1619–2003 AD) (Nagovitsyn, 2006) *(http://www.gao.spb.ru/database/esai).*
5. The Meudon filament data series (1919–1991 AD). These data are taken from "*Cartes Synoptiques de la Chromosphere Solaire et Catalogues des Filaments et des Centres d'Activity: 1919—1989, published by Observatoire de Paris, Section de Meudon*" and for the period 1989-1991 from *http://bass2000.obspm.fr/lastsynmap.php*.

The first series (*GSN*) is taken as a better proxy of the solar short wavelength electromagnetic radiation and the corpuscular fluxes variability than the classical *Ri* –index before the middle of 19$^{th}$ century. Unfortunately, it ends at 1995 AD and there are no plains for its continuation at this stage (Kane, 2008a).

The extended simulated sunspot series (*Rsi*) is the longest used in the present analysis – 913 years. The *Rsi*-series is based on the "historical" Schove's series (Schove, 1983) for the moments of extremes and magnitudes of the Schwabe-Wolf' sunspot cycles. The mean annual sunspot numbers there have been derived on the basis of Krylov-Bogolyubov's approach to the description of weakly nonlinear oscillatory processes (Nagovitsyn, 1997; Nagovitsyn *et al.*, 2004). It is important to note that the original Schove's series is based predominantly on historical messages for auroral events and naked eye visible sunspots (i.e. potential sources of strong flares). By this reason it could be considered as a good proxy of the solar flare activity and its active centers.

The synthetic type of this series lead to the reasonable question about the significant errors in these data, which may be exist there. The problem should be significant mainly for the pre-instrumental part of the *Rsi*- series, i.e. before 1610 AD. Concerning that one it should be refer to the error estimations of Nagovitsyn (1997). He found that the error range of the corresponding series is ±30% to the mean annual *Rsi* data. However, this error should be decreased essentially by smoothing of the data over 11 years (see below). The reliability of the

ESAI sunspots (i.e. *Rsi*-series) is also tested by Nagovitsyn (1997) due comparison with other "historical" solar activity data sets ($^{14}$C, the naked eye visible sunspots and the last Schove's series version (Schove,1983)).

We have provided some additional comparisons between the Schove's series and Rsi for the epoch 1090-1610 AD. The typical uncertainty of the 11-year sunspot cycles magnitudes in Schove's series is 25-30% and it is comparable with the same one for *Rsi* mean annual values. The both series has been also compared with standpoint of validations or violations of the "amplitudal" Gnevishev-Ohl's rule. It has been found that for the all 24 pairs of even-odd 11-yaer cycles during the epoch 1090-1610 AD, there is coincidence for validation (or violation) in 19, i.e. for ~80% of the cases. In four cases from the other five ones the corresponding pairs cycles are very weak by amplitudes - they has been occurred during the Wolf 's and Spoerer's minima in 13$^{th}$-14$^{th}$ and 15$^{th}$ century, respectively. Thus, one could conclude that the both series are in good agreement each to other in almost all important aspects.

The length of the simulated *aa*-index (*AA*) is almost the same as the *GSN*. The Meudon filament series is relatively short and it is analyzed only with the standard T-R periodogram and wavelet analysis. That series is not long enough for studying the evolution of cycles longer than the Schwabe-Wolf's cycle. However, it is included here as a proxy of the long-lived solar magnetic field structures.

Finally, the Zurich sunspot *Ri* annual series is used as an international standard for the overall sunspot activity.

Two preliminary procedures have been applied to all the studied time series:
1. Removing of general nonlinear trends (polynomials of second, third or fourth degree). The corresponding trend functions were obtained by the means of a least mean squares procedure. The best trend function expressions were determined on the basis of the best coefficients of correlations to the corresponding time series and the Snedekor-Fisher's *F*-parameter.
2. A smoothing procedure by 11 points (years) over the "residuals" was applied. The effect of Schwabe-Wolf's cycle has been removed and the signal of the long term cycles is much better expressed.

We apply three type of time series analysis over these data sets.

A. The standard T-R periodogram procedure

The standard T-R periodogram analysis is used for searching for statistically significant cycles of the whole time series (Komitov, 1986, 1997; Benson *et al.*, 2003 etc.). This technique is very close to the algorithm, described by Scargle (1982). A more detail description in the Komitov's paper (1997) is given.

The idea of this method is to approximate the studied time series *F(t)* by minimized function $\varphi(t)$ of simple periodic type, i.e.:

$$F(t) \approx \varphi(t) = Ao + A\cos(2\pi t/T) + B\sin(2\pi t/T) \quad (1)$$

where $t = 0,1,2…$ are the corresponding moments in time step units (the time series step). *Ao* is the mean value of *F(t)* based on the entire time series. *T* is the period, which is varied in the range [$T_o$, $T_{max}$] by a step of $\Delta T$. This way, a series of *p* minimized functions $\varphi(t)$ is obtained, where $p=(T_{max} - T_o)/\Delta T$. The minimal possible value of $T_o$ is equal to 2 steps of the time series ($\nu=0.5$). For each one of the so obtained functions $\varphi(t)$ a correlation coefficient *R* to the time series *F(t)* is calculated. The local maxima of *R* point out the possible existence of cycles with duration equal to the corresponding periods *T*.

The spectral resolution step used in the standard T-R periodogram calculations for the time series in this study is $\Delta T=0.5$ years. *T* varies from 2 to 502 years, i.e. the T-R correlograms are calculated for 1000 different values of *T*.

B. "Moving Window" T-R Periodogram Procedure (MWTRPP)

The standard T-R periodogram procedure produces the mean parameters of the existing cycles in the time series. However, these cycles could significantly change between different parts of the time series. For studying a cycle's evolution the so called "Moving Window T-R periodogram procedure" (MWTRPP) (Bonev *et al.*, 2004) is used. In this algorithm, a part of the time series of length $P$ ("moving window"), where $P < N$ is defined. By author's opinion it is recommended that $P / N \leq 1/3$ (Komitov, 2009). At the start of the procedure the "moving window" contains the first $P$ terms of time series $F(t)$ and the standard T-R procedure is applied to those. After that the "moving window" is shifted with a step $\delta T$ (one or more integer time series step), and the T-R procedure is repeated again, etc. Using this method, a series of T-R correlograms could be obtained, in which the y-coordinates corresponding to $T$ values are presented as columns in two-dimensional maps, while the x-coordinates correspond to the central or starting moments of "moving window" epochs. In addition to $R$ values, one could obtain maps of $R/SR$ where $SR$ is the error of $R$, the amplitude $a(T)=\sqrt{(A^2(T)+B^2(T))}$ as well as the evolution of the coefficients $A(T)$ or $B(T)$.

C. The wavelet- analysis

This technique is well known. It has been extensively applied in complex nonlinear time-series analysis, including the study of solar and stellar activity cycles (Torrence and Compo, 1998; Sello 2003, Ranucci and Sello, 2004 etc.).

*3. Results and analysis*

*3.1 The Zurich series (1750-2009 AD)*

As it is shown on Figure 1, where the T-R correlogram is plotted, the main long-term cycle in the international sunspot number $Ri$ series is approximately 95.5 years long. The corresponding peak of the correlation coefficient $R$ exceeds its error 18.8 times. The "zero-hypothesis" probability in this case is $<< 10^{-6}$. Consequently, the quasi-centurial cycle should be considered as the most important feature of the international sunspot index $Ri$, after the 11-year Schwabe- Wolf's cycle. The second clearly visible long-term oscillation has T= 58.5 years. The corresponding $R/SR$ ratio is also very high (~8.8) and with significance > 99.999%, i.e. "zero-hypothesis" level is less than 0.001%. There are also traces of 41-year and 29-year oscillations that are weaker, but with high statistical significance.

The MWTRPP amplitude map of the Zurich series is shown on Figure 2. An evolution of the quasi-century oscillation during the last 260 years is clearly visible. During the first 150 years, i.e. the second half of the 18th and the 19th centuries this cycle is noticably shorter (~70 years). However, during the next decades its duration is slowly increased and in the 20th century it is already slightly longer than 100 years. The calendar centres of the moving window epochs when the quasi-centurial cycle is better expressed are near to 1825 AD (the Dalton minimum) and 1870 AD.

There is no good trace of a ~55-year cycle during the 18th century. As it is shown, this cycle is in process of "separation " from the quasi century oscillation during this time. The ~55-year cycle is well expressed in the 19th century up to 1870 AD, but after that it is decreasing both in amplitude and duration. A new increase in the amplitude is observed in the 20th century, but now the cycle length is about 40-42 years. On other hand, a quasi 42-44 years cycle is very well expressed in the middle of the 19th century almost simultaneously with the 55-year one. Another visible weak oscillation is the quasi 30-year cycle (~3 Schwabe-Wolf's cycles). Its amplitude has increased and decreased three times during the whole period of 260 years. The last one corresponds to the recent decades.

The results from the wavelet-analysis (WA) are shown in Figure 3. The strongest cycle here has a duration of ~ 94 years. It is slightly shorter during the 18th century with a weak tendency of prolongation during the 19th and 20th centuries. In the recent part of the series (20th century) this cycle tends to a duration of ~100-110 years. The second by importance cycle is the 54-year one. It is very well expressed in 18th and 19th centuries. After ~ AD 1850 its amplitude is fading.

### 3.2 The Group sunspot number series (1610-1995 AD)

According the results from the T-R periodogram analysis the main cycle in the *GSN* series during the last ~ 400 years is with quasi-two-century ~ 202 years duration (Figure 4). The corresponding coefficient of correlation *R* is ~0.58. The second by significance (*R*= 0.52) is the 108-year cycle. There is also a quasi 80 years oscillation. The sub-century 54-year cycle is the next significant oscillation (*R*= 0.33). There are also very weak traces of 28- and 21.5-year cycles. The ratio *R/SR* for the last two ones is in the range of 2.0 to 3.5.

On the basis of MWTRPP–method we found that the quasi-two-century oscillation was very powerful during the $17^{th}$ and $18^{th}$ centuries, i.e. during the Maunder minimum and the first decades after that. However, since the Maunder minimum it quickly fades and is transformed to shorter duration. Close to the Dalton minimum (1795-1830 AD) there are no more visible cyclic tendencies of a century or longer duration. The main long term cycle at the end of the Dalton minimum has a period of about 70-75 years. After the Dalton minimum, a slow prolongation starts for this cycle and its "actual" length during the $20^{th}$ century is approximately 110 years.

A relatively stable 50-60 years cycle is clearly visible in Figure 5 for almost the entire GSN data series. This cycle is best expressed in the $18^{th}$ and $19^{th}$ centuries up to 1870 AD, and in the $20^{th}$ century it is significantly weaker. During the last decades a weak quasi 30 year cycle appeared.

According to the wavelet analysis the quasi ~200 year cycle is most powerful in the earlier part of the GSN series (Figure 6) Its amplitude decreases slightly from the $17^{th}$ to the $20^{th}$ centuries. In contrast, a cycle with duration of ~110 year increases in amplitude during this time. Another ~80-year oscillation, which exists during the $17^{th}$ and $18^{th}$ centuries quickly converges to the 110-year one approximately after the Dalton minimum. A 54-year cycle exists in the whole series, but it is most powerful during the $18^{th}$ and $19^{th}$ centuries. The peak of its amplitude is near 1820 AD. As in the MWTRPP-map (Figure 5) a 30-33 years cycle is well visible in the most recent part of the GSN-series.

As in the case of *Ri* (Zurich series), the wavelet amplitude spectra of *GSN* (Figure 6, right) is very similar to the corresponding T-R correlogram (Figure 4).

*3.3 The simulated AA-index series (1619-1999 AD)*

The length of the simulated *aa*-data series from ESAI used here is almost the same as that of the GSN–series (381 years). The signature "*AA*" is used in this work to distinguish it from the instrumental *aa*-index . The study of this series is very interesting because of the possibility for comparison of the results of this purely "geophysically oriented" series to the other ones , which are related to the sunspot active centers. The standard T-R spectrum is shown in Figure 7.

According to MWTRPP the quasi 55-60 years cycle is very stable in the earlier and middle part of the series (Figure 8) and it is slightly better expressed in the *AA* than in the *GSN* and *Rsi* series. On other hand the WA results reveal that this cycle is even better expressed during the $17^{th}$ century in *AA* (Figure 9) than in the cases of *GSN* and *Rsi*.

There is a hint of a better expressed quasi 30-year oscillation during the earlier part of the *AA*-series (Figures 8 and 9). According to the WA the higher amplitudes of this cycle are at the end of the $20^{th}$ century (Figure 9), while as follows from MWTRPP, the absolute amplitude peak of the quasi 30-year oscillation occurs almost a century earlier (Figure 8).

*3.4 The Meudon filament data series (1919-1991 AD)*

Unlike the other data sets investigated here, the smoothed filament series of the Meudon observatory catalog is relatively short, covering only 73 years . For this reason, it is not possible to effectively search it for any cycle evolution in the multidecadial range. Therefore, we study only the mean features of the whole series and mainly on the basis of the standard T-R periodogram analysis (Figure 10). As it is shown, there are 66.5-, 26.5- and 17.5-year cycles. The WA-test gives, due to the shortness of this time series, only a weak 26.5-year cycle (Figure 11).

These results are interesting due to the fact that the filaments are a relative "pure" indicator for the coronal activity phenomena. The existence of 66.5-year cycle is evidence that the subcentury periodicity is real for these events at least during the 20$^{th}$ century. The absence of quasi 117-year cycle, which was detected earlier by Duchlev (2002) could easily be explained by the de-trending procedure.

*3.5 The long term solar cycles during the last ~900 years*

The time series used in Sections 3.1-3.5 are relatively short. They all begin near, or after the supermillenial Maunder minimum (1640-1720 AD). Moreover, they are almost entirely contained within a period of a long term upward trend of the solar activity, during the initial active phase of the quasi–bimillenial solar 2200-2400 year cycle (Hallstadtzeit ) (Damon and Sonett, 1991; Dergachev and Chistyakov, 1993; Bonev et al., 2004). As it has been already demonstrated by some of these authors (Damon and Sonett, 1991; Bonev *et al*., 2004) on the basis of $^{14}$C data, as well as by Komitov *et al*. (2004)( $^{14}$C and Schove's series), Maunder-type minima are not only the starting phases of Hallstadtzeit cycles. Serious changes of the solar activity dynamics occur during these epochs. The amplitudes of the quasi two-century cycles (~170-220 years) fade, while in their place an increase of the cycles by quiasi- century duration begins. This suggests that it is interesting to search for the stability and evolution of the cycles from the studied range over longer than 400 years time scales. It is especially interesting to investigate how the transition from the previous Hallstadtzeit cycle to the present one affects the solar oscillations with sub-century periods (20-70 years).

For this purpose, we use the whole *Rsi* simulated data series from the ESAI. As it has been already noted in Section 2, this series starts from 1090 to 2002 AD and contains annual data for 913 years. The T-R periodogram analysis is used in its both standard and MWTRPP versions. The "moving window" length *P* is 400 years. All other parameters are the same as those described in Section 2. As for the other series, the trend is removed and an 11 years smoothing procedure has been performed. The larger width of the window *P* provides much better conditions for the MWTRPP than in Sections 3.1-3.3 over all and especially in the range of T ≥ 150 years. The results are shown on Figures 12, 13 and 14.

The main cyclic oscillation in sunspot activity during the last ~900 years is with duration of 204 to 209 years (Figures 13 and 14). According to Figure 12 there is a ~121.5-year cycle and an oscillation with duration of 82 years, i.e. very close to the so-called "Gleissberg cycle" (78 years)(Glessberg, 1944). There is also a very well expressed 59.5-year cycle. A weaker cyclic component at T=54 years is also present. Weak signatures of 41- and 29-year cycles are found, which are near or even less than the critical level *R/SR*= 3.46.

As it is shown on Figure 13 the quasi 200-year cycle is very significant before the Maunder minimum epoch. The absolute maximum of its amplitude corresponds to the calendar center of the 400 years window at ~1550 AD. This epoch contains both of the deepest solar supercentury minima during the last 2000 years - those of Spörer (1400-1520 AD) and Maunder (AD 1640-1720). This result matches very well all other previous evidence, that during the Hallstadtzeit cycles minim, the amplitude of the quasi two-century cycle tends to a maximum, while the quasi-century ones – to their absolute amplitude minima (Damon and Sonett, 1991; Bonev *et al*., 2004; Komitov, 2007). According to the WA results (Figure 14) the amplitude of the ~200 yr cycle tends to reach a maximum near the Spörer–Maunder minimums epoch , but the fading after that is less pronounced than in Figure 13.

The relative fading near the Maunder minimum (moving window calendar center at ~1600 AD) of all cycles by duration 40-140 years is well pronounced in Figure 14. But it should be noted that in the entire period of 11$^{th}$- 20$^{th}$ centuries, three very stable oscillations in the sub-century and quasi century range are always present, corresponding to mean periods of 55-60, ~80 and ~ 120 years. They could be followed in Figure 13 during the entire ~900-year period. One could conclude that there is no general mean quasi-century cycle during the last millenium, but rather a quasi century doublet ( $T$= 80 and 110-120 years) or triplet if the subcentury 50-60 years cycle is also considered.

The amplitude of the longest ~120-year component reaches local maxima near the 11$^{th}$, 15$^{th}$ and 20$^{th}$ centuries. The behavior of the ~55-60 years cycle is also interesting. The epochs of its higher amplitudes are after the Maunder minimum, as well as in 11$^{th}$-12$^{th}$ centuries. Between the 12$^{th}$-17$^{th}$ centuries the amplitude of the 55-60 years oscillation remains significant, but not so high.

It is shown on Figures 13 and 14 that unlike the quasi-century multiplet ( 50-60, 80 and 120 years) the shorter cyclic variations are not stable on longer time scales. There are only weak traces of the 40-45 year cycles predominantly before the Maunder minimum and during the 20$^{th}$ century (Figure 13). Traces of the ~30-year (~3 Schwabe-Wolf's cycles) oscillation are even weaker and sporadic. As can be seen, the ~30-year cycle is slightly better defined during the latest part of the series and mainly due to its significant increase in amplitude during the 20$^{th}$ century (Figure 13 and 14).

It is interesting to note that near the Maunder minimum, significant traces of ~35-38 years cycle are visible in Figure 13 (the MWTRPP map). An oscillation with such duration has been detected for the period of 1932-2005 AD in the annual number of the geomagnetic storms, when the *Ap* –index exceeds 40 ( Komitov, 2008).

*4. Discussion*

We could discuss the presented there results and their analysis in two main aspects: 1. The long term cyclic solar variations and their importance for solar activity predictions in short and long time scales at all; 2. The long term flare activity dynamics.

*4.1. The long term solar cycles and deep solar minima predictions*

The results presented in Section 3 provide clear evidences that there are three significant and relatively stable quasi-oscillations during the last millenium with sub- and quasi-century duration. Their durations are of 55-60, ~80 and 110-120 years, respectively. As it has been shown in Sections 3.1-3.4 both by the WA and the MWTRPP methods their periods are slightly variable in the shorter time scales, for example the epoch of instrumental observations, i.e. the last ~400 years. On other hand, the MWTRPP test over the entire $Rsi$ series shows that the reliable presence of this quasi-century multiplet covers the whole ~900-year period if a larger smoothing window is used.

However, as is shown in Figures 2, 3, 5, 6, 8, 9, and 14, the ratios between the amplitudes of these three oscillations are different in the different epochs. This could explain why there are serious variations in the length of the observed quasi-century cycle in the different epochs. Gleissberg (1944), found a length of ~78-80 years. In his study, the data used referred to sunspot cycles No 0-17, when the ~110-120 year component is noticeably weaker. Thus, the ~94-95 years century cycle in the whole Zurich series (Figures 1 and 3) is only a "mean-weight" one and it is an integral result of the more important role of the 55-80 years and the ~80-year cyclic components in the earlier part of this series and the fading and increasing of the ~120-year component in the recent part of the series.

On the other hand, the 50-60 years cycle is strong only in the first half of this period (before ~1870 AD), but it is interrupted after the Gleissberg-Gnevishev minimum epoch (1898-1923 AD). However, the strengthening of the ~120-year component and the weakening of the 80-year one after the Dalton minimum and during the 20$^{th}$ century is the cause for the delay of the next long-term minimum. That minimum began not after the end of solar cycle No 21 in 1986 AD, but about 20-22 years later, i.e. at the end of cycle No 23. The imminent long-term minimum in the first half of the 21$^{st}$ century has been predicted by many authors (Fyodorov *et al*, 1995; Badalyan *et al*., 2000; Komitov and Bonev, 2001; Komitov and Kaftan, 2003; Shatten and Tobiska, 2003; Solanki *et al*.,2004; Ogurtsov, 2005; Cliverd *et al*.,2006). As a consequence of that, the amplitude of solar cycle 24 is expected to be relatively low ($Ri_{max}$ ~ 90) according Pesnell *et al*., (2009) or even ~58 (Kane,2010)

The complicated structure of the quasi-century cycle was established as early as Schove (1955) on the basis of historical records of auroral events for the last ~2000 years. In addition to the Gleissberg (~78 years) cycle, he found that there are also traces of longer (~120-130 years) or shorter (54-55 or 65 years) oscillations.

Nagovitsyn (1997) found that there should be two "fundamental" quasi-centurial cyclic components in the "extended" Wolf's series (1700-1990 AD) with duration of 80 and 115 years, respectively. This is in very good agreement with our results there. However, we found also ~60-year cycle and that is confirmed by two quite independent methods (TRPA and WA). How could be explain its existence, as well as its high stability?

The existence of ~60-year cycle could be explained with the fact that this cycle is in high degree in resonant correlation to the Hale 20-22-year magnetic cycle, 11-year Schwabe-Wolf cycle, as well as to the 120 yr cycle. Three Hale cycles are ~60-65 years, five Schwabe-Wolf cycles are ~50-55 years, six Schwabe-Wolf cycles are ~ 60- 66 years, and 120-yaer cycle is divisible by 60. The Hale and Schwabe-Wolf cycles are the main ones in the solar dynamo action. It is interesting, in this context, that in the long $Rs$i-series the standard T-R periodogram analysis detects two close adjacent components ( 54.5 and 59.5 years, see Figure 12). This indicates that all of the above-mentioned resonances should play role rather than only one or two by them. However, by our opinion it is difficult to explain the stability of the 60-year cycle only on the base of these complicate resonance correlations. Most probably there should be an additional independent source of these oscillations, which give a contribution to the overall solar magnetic flux variability. The period of its variations is ~55-60 years and occasional it is approximately in resonance with the 10-11 year Schwabe-Wolf, 20-22 year Hale, and 120-year cycles.

There are some other interesting studies, concerning the existence of solar cycles with duration of quasi–3 (Ahuwalia, 1998) and quasi-5 (Du, 2006) Schwabe-Wolf cycles, i.e. 30-33 (three cycle periodicity, the so called "TRC-rule" ) and ~ 55 years, respectively. The existence and stability especially of the TRC has been analyzed critically by Kane (2008b). He has found that there are only three sequences of Schwabe-Wolf cycles during the last 300 years, for which the TRC-rule is valid and two of them covered the Zurich cycles No 17-22. On the other hand, the quasi-4 Schwabe-Wolf's cycle periodicity (~40-45 years) has been detected in the north-south asymmetry of the sunspot area by Javaraiah (2008) and Komitov (2010).

In the present study, we found that the ~30 and 40-45 years oscillations are present in separate epochs in the overall sunspot and geomagnetic activity, but in contrast to the century multiplet components, they are very unstable and weak. However, both cycles have significantly higher amplitude in the recent epoch. The 40-45 year cycle is in resonant correspondence to the Hale cycle, while the same could not say for the 29-30 yr one.

It is necessary to note that there is no strong correspondence between the observed ~30-year cycle and the "TRC-rule". The ~30-year cycle in our study is a feature of the absolute amplitude variations of the smoothed and de-trended series of solar indices, while the "TRC-rule" describes a relative relationship between three consecutive 11-year Schwabe-Wolf's cycles.

As we have already noted, the ~200-year cycle is one of the main features of the solar activity on the supercentury time scale. It is very clear visible in "cosmogenic" radioisothopes data (the so called "de Vries oscillation" in $^{14}$C series) )( de Vries, 1958; Stuiver and Quay, 1980; Damon and Sonett, 1991; Dergachev and Chistyakov, 1993). We found here that this cycle has been very powerful before the Maunder minimum and obviously weaker after that. Its decreasing amplitude during the last ~400 years of instrumental observations is detectable, but it is quite gradual according to the WA-method. However, this fading is abrupt according the MWTRPP procedure. This is caused by the use of relatively narrow "moving window" ($P=150$ years) for studying the shorter series (Figures 2, 3, 5, 6, 8, and 9), which makes the method not precise enough for cycles of the same or longer duration. On the other hand, an additional disturbance over the ~200-year cycle is caused by the Gleissberg-Gnevishev's minimum (1898-1923 AD). The detected dynamics of the 200-year cycle match previous results and conclusions about its amplitude modulation by the Hallstadtzeit cycle (Damon and Sonett, 1991; Bonev *et al*., 2004; Komitov *et al*.,2004).

The results presented in this study are an additional confirmation for the important role of the long term cycles for the solar activity dynamics. The high relative stability of the quasi-centurial multiplet (~60, 80 and 115-120 yr), as well as of the quasi- bicenturial (~200 yr) cycle is an evidence, that they are much rather regular than stochastic phenomena. As it is pointed in many studies last one is valid for the Maunder-type minima, too (the 2200-2400 Hallstadtzeit cycle). It is very probably that some of them are caused by long term cyclic processes that operate in the deeper layers of the solar convective zone. Their physical nature is quite different from the standard "solar dynamo" model and related to it 11 and 22 yr oscillations. The question for the nature of these processes (solar diameter changes, differential solar rotation variations etc. see Sokoloff (2004) and references therein)), how they are connected to the convective zone transport and to solar dynamo phenomena is unclear yet. It is strong necessary to take into account these long term features of solar activity for its more reliable predictions. Their effects should be used in more precise physical models of solar variability because the present "solar dynamo" models give an account only the 11 and 22 yr cyclic evolution of the solar magnetic field.

An additional study especially focussed over the existence, stability and evolution of the subcenturial solar cycles is necessary. We plane to use the long "cosmogenic" $^{14}$C time series with relative high time resolution INTCAL04 (Reimer et al., 2004). It should be also a very important independent test for the validity of the presented there results for the last millennia, based on the synthetic Rsi- series.

*4.2. The long term flare activity dynamics*

It is very probable that the three distinct components of the quasi-century multiplet are connected to the long-term dynamics of different classes of active centers and sources of flare activity. On the basis of Křivský and Pejml (1988) catalogue data Komitov (2009) found that a strong 62.5-year cycle exists in the annual numbers of middle latitude aurora (MLA) events during the 18th and 19th centuries. In this study, they also find that the peaks of this auroral cycle correspond very well to the quasi 60-year cycle maxima of $^{10}$Be production rates in the "Greenland" beryllium-10 data series (Beer *et al*., 1990, 1998).

It is interesting to note three additional facts here, namely: 1. The 11-year cycle is weak in the T-R spectra of MLA annual number (Komitov 2009), while the 62.5-year one is very strong (the corresponding peak in the coefficient of correlation *R* is ~0.65). This result was found without the application of any smoothing procedure. 2. There are closely coinciding peaks both in the MLA and the $^{10}$Be production rates near 1725-1735 AD, ~1800-1805 AD, and 1865-1870 AD (there is a slight delay in $^{10}$Be in the range of 3-4 years, which could be particulary explained by the "resident time" (Damon *et al*., 1997)). 3. The annual number of MLA events has drastically decreased after 1870 AD (Figure 18) when the 50-60 years cycle in *Ri*, *GSN*, *AA,* and *Rsi* is weaker (Figures 2, 3, 5, 6, 8, and 9).

The strong quasi -60 yr cycle in MLA events annual numbers ($N_{MLA}$) during the period of 1700-1900 AD is very clearly shown in Figure 15 (the lowest panel). There are three local main maxima at 1730, 1787-88 and 1850 AD. There is also another significant maximum at 1870 AD. The $N_{MLA}$ dynamics is quite different from those of the *AA*-index (the middle panel) and *GSN* (the upper panel). All these MLA maxima are related to corresponding sunspot Schwabe—Wolf's cycles ones (in 1730 AD to SC-2, in 1787-88 AD to SC4, in 1850 AD to SC9 and in 1870 AD to SC11). However, as it is shown, in the short –time structure of $N_{MLA}$ series the 11-year cycle is not clear at all. It is seriously damaged by many other low amplitude variations on order of a few years. The long term changes of GSN and $N_{MLA}$ are also not well matched. For example there is a general increase in the sunspot cycle amplitudes between 1730 and 1770 AD , while the MLA activity decrease during this period. Another period of large differences occurs between 1870 and 1900 AD. There is a relatively high sunspot maximum No 13 in 1890th , but the MLA event are very rare during this period. The most powerful sunspot cycle during the 19th century is SC8, but the higest MLA activity is during the maximum of the next SC9. So, there is no strong relationship between the sunspot activity and the MLA events. It is possible that during strong sunspot Schwabe-Wolf's cycles the corresponding MLA activity could be low and high MLA activity could occur during weak sunspot cycles. The coefficient of linear correlation between GSN and $N_{MLA}$ is *r*= +0.57 for the epoch 1700-1900 AD. It follows by the Snedekor-

Fisher's *F*-test (in this case *F*=1.48, the ratio between the total and residual variances) that only 32% of the MLA activity variations could be directly related to the overall sunspot activity changes.

The situation is almost identical if the geomagnetic activity (*AA*) and the $N_{MLA}$ are compared. In this case *r* =+0.61. Consequently, there is not a very close relationship between the geomagnetic activity and MLA either. It should also be noted that *r*= + 0.73 (*F*=2.17) between *GSN* and *AA* during the same epoch (1700-1900 AD).

The MLA phenomena are usually associated with strong solar flare events, which also cause coronal mass ejections (CME). Thus, taking into account the afore-mentioned three facts it could imply that the 50-60 years cycle is a feature of these active solar centers, which are the typical sources of strong flares and CMEs. The coincidence of the local peaks in the 60-year cycles of $^{10}$Be and the MLA during the 18$^{th}$ and 19$^{th}$ centuries suggests that a significant fraction of the $^{10}$Be atoms could be produced in the stratosphere by highly energetic particles from strong solar flares (Usoskin *et al.*, 2006). This could make the relationship between the sunspot activity and $^{10}$Be production rates much more complicated on century or subcentury time scales compared to the case of only a pure Forbush- effect (Komitov, 2009).

There could be another point of view on the possible relation of the 50-60 years cycle to the sources of the strongest solar flares. It concerns the historical data for giant, naked eye visible sunspots and sunspot groups. As was shown by Vaquero (2007), there are three peaks of the 50 years smoothing annual numbers of the giant sunspots during the 18$^{th}$ and 19$^{th}$ centuries, near 1735 AD, 1805 AD, and 1870 AD. These three peaks correspond well to local maxima of the MLA annual number (Figure 15) and the "Greenland" $^{10}$Be concentrations data. The strong downward trend in the annual number of giant sunspots after 1875 AD coincides well both with the corresponding downward tendency of MLA (Komitov, 2009) and the amplitude of the 55-60 years oscillation (Figures 2, 3, 5, 6, 8, and 9 in Section 3). Thus, it is very probable that the 55-60 years cycle in the sunspot activity is connected directly to the active regions, which are the sources of the strongest solar flares, corresponding to the X-ray classes M and X.

Vaquero (2007) shows that there is no significant correlation between the *GSN* (i.e. the overall sunspot activity) and the visible by naked eye sunspots at all. On the contrary, there is a significant anticorrelation or non-correlation over large fractions of this period. It is also evident from Figures 13 and 14, that the amplitude changes of the 55-60 years cycle are the most independent relative to the two other components of the quasi-century multiplet. Let us also take into account the very small amplitude of the Schwabe Wolf's cycle in the MLA events (Komitov, 2009). By combining and comparing all these facts, we arrive at the conclusion that these active centers, which are related to the most powerful flare classes and are subject to the 55-60 years cycle are relatively independent of the overall sunspot activity dynamics. Thus, the weak correlation between X and M class flares and the overall sunspot activity *Ri* during the last 35-40 years (Table 1) is a consequence of the same weak relationship on long time scales.

Table 1.The coefficients of linear correlation between the monthly values of 6 solar and solar-modulated indices for the period January 1980 - December 2009

|  | *Ri* | *F107* | *Nflux* | $N_C$ | $N_M$ | $N_X$ |
|---|---|---|---|---|---|---|
| *Ri* | 1 | 0.975 | -0.798 | 0.847 | 0.719 | 0.449 |
| *F107* | 0.975 | 1 | -0.796 | 0.856 | 0.759 | 0.487 |
| *Nflux* | -0.798 | -0.796 | 1 | -0.724 | -0.615 | -0.464 |
| $N_C$ | 0.847 | 0.856 | -0.724 | 1 | 0.672 | 0.415 |
| $N_M$ | 0.719 | 0.759 | -0.615 | 0.672 | 1 | 0.770 |
| $N_X$ | 0.449 | 0.487 | -0.464 | 0.415 | 0.770 | 1 |

*Ri* - the international sunspot number
*F107* - solar radioflux at *f*=2800 MHz (λ=10.7 cm)
*Nflux* – GCR neutron flux (Moscow)
$N_C$, $N_M$ and $N_X$ - monthly numbers of X-ray flares of C, M and X classes by GOES satellite dat

*5. Conclusions*

By applying the essentially independent methods of the T-R periodogram analysis, both in the standard and "moving window" regime, and the wavelet analysis on five different solar and geomagnetic data sets a detailed analysis of the problem of the existence and stability of cycles with duration in the range of 20 – 250 years was made. On the basis of the results obtained in this study, the following main conclusions can be made:

1. The quasi century cycle in sunspot and geomagnetic activity has a complicated multiplet structure, which is clearly detected in different types of instrumental and simulated indirect data on supercentury time scales ( ~400 years or longer). There are three important and relatively stable components of this multiplet with durations of 55-60, ~80 and ~120 years, respectively. Most probably they are related to different types of solar active centers and sunspot groups. The Gleissberg's ~78-80 years cycle is only one of these components.

2. There is a tendency during the last ~150 years towards increasing the amplitude of the quasi 120-year cyclic component and fading of the ~80-year one. There is an indication that a ~400 yr cyclic amplitude modulation of the 120 yr component exist.

3. It is very probable that the quasi 50-60 years cycle is related to the sources of the most powerful solar flare events (x-ray classes M and even more X) and CMEs. They are relatively independent from the overall sunspot activity dynamics and the 11- year Schwabe-Wolf's cycle. Their relative participation in the overall solar and geomagnetic indices is low.

4. The quasi 40-45 and 30 years cycles in the solar and geomagnetic activity indices are unstable on long time scales and their amplitudes are low. A visible strengthening of the quasi 30-year periodicity during the last few decades is observed.

**Acknowledgements**

The authors are grateful to Prof. Yu. Nagovitsyn for his help with compiling the data from ESAI (*http://www.gao.spb.ru/database/esai*). The authors are thankful to the National Geophysical Data Center, Boulder, Colorado, U.S.A. for providing the annual Group sunspots number and international sunspot number data via *ftp://ftp.ngdc.noaa.gov/STP*, and the Meudon Observatory, Paris, France for providing the filament number data via *http://bass2000.obspm.fr*. The authors are thankful also to Prof. I. Usoskin for important comments, concerning some of the results and their analysis.

Institute of Astronomy, 72 Tsarigradsko Chaussee, Sofia 1784, Bulgaria
b_komitov@sz.inetg.bg
duchlev@astro.bas.bg
mdechev@astro.bas.bg
koleva@astro.bas.bg

**Stefano Sello**
Mathematical and Physical Models, Enel Research, Via Andrea Pisano 120, 56122 Pisa - Italy

**Kaloyan Penev**
Harvard University, Mail Stop 10, 60 Garden Street, Cambridge, MA 02138
e-mail: kpenev@cfa.harvard.edu

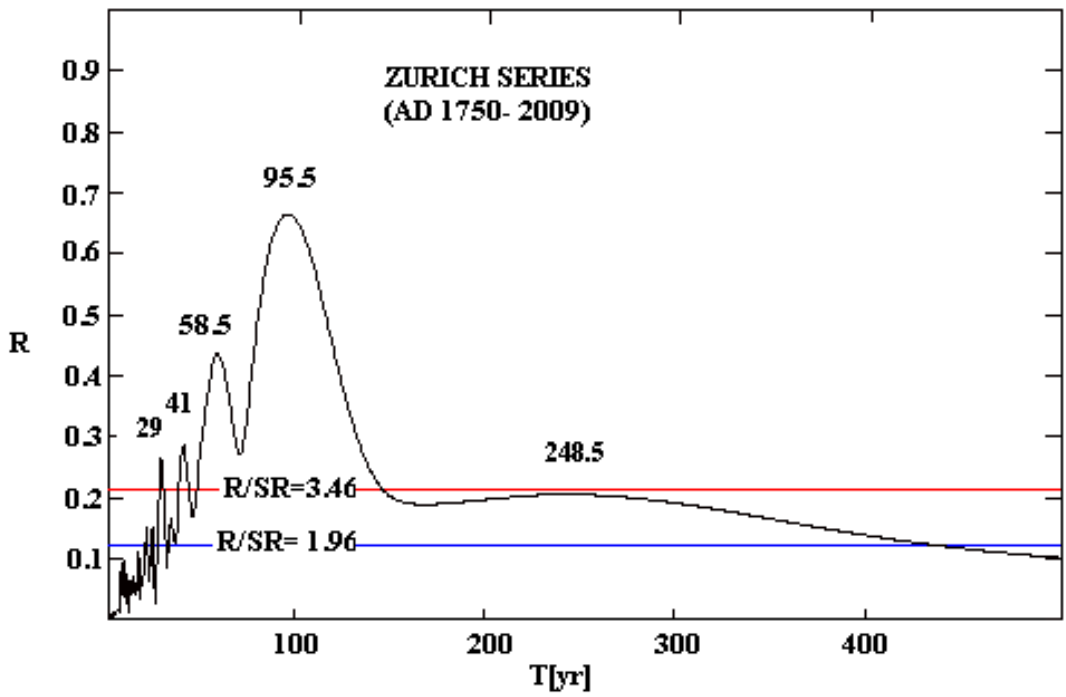

Fig. 01 T-R corellogram of the Zurich series (AD 1750-2009)

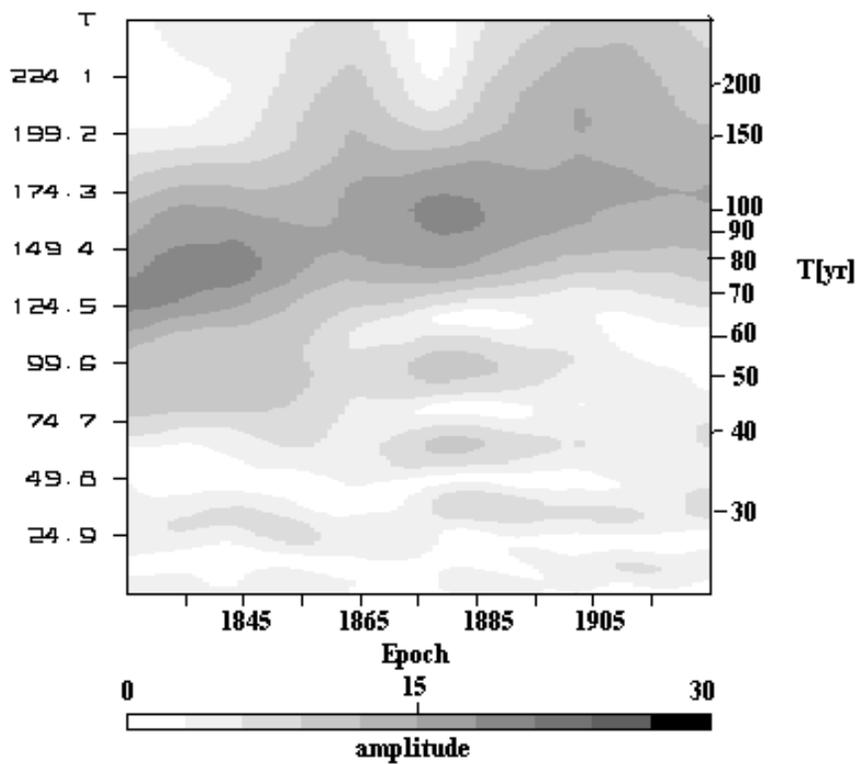

Fig. 02 MWTRPP amplitude map of the Zurich series (AD 1750-2009)

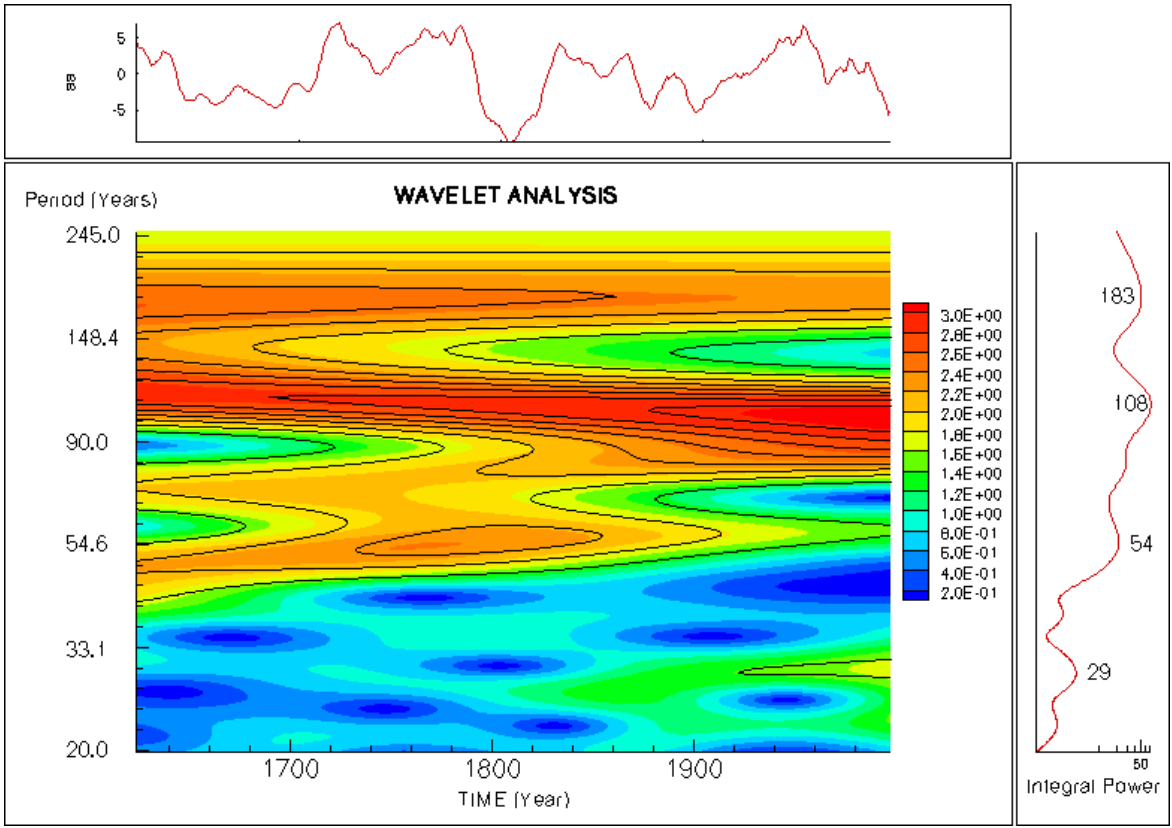

**Fig. 03 WA amplitudes of the Zurich series ( AD 1750-2009)**

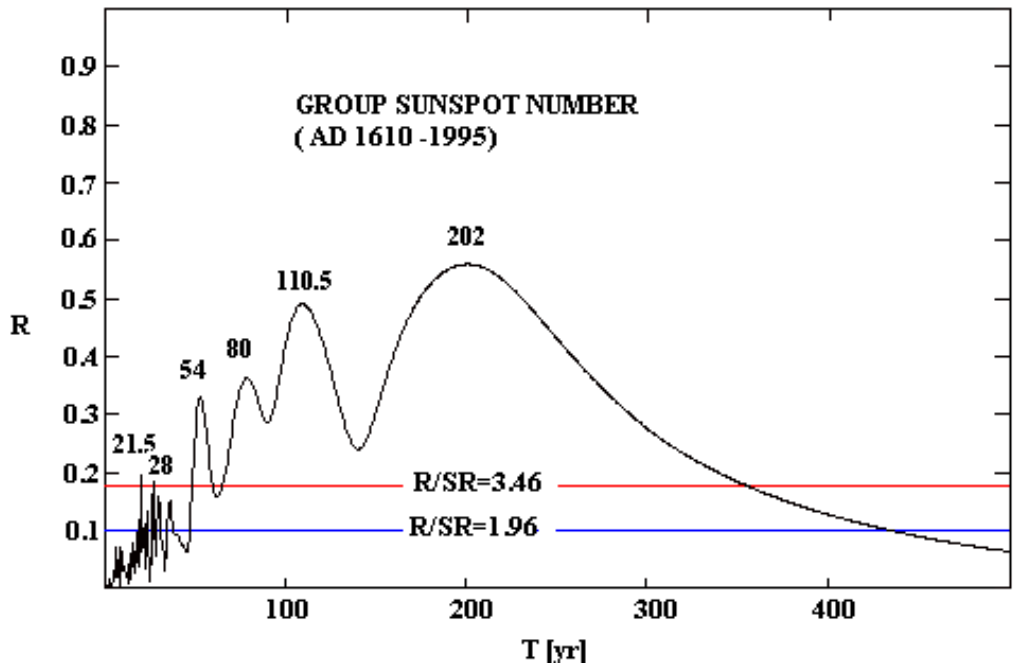

**Fig. 04 T-R corellogram of the Group sunspot number (*GSN*) series (AD 1610-1995)**

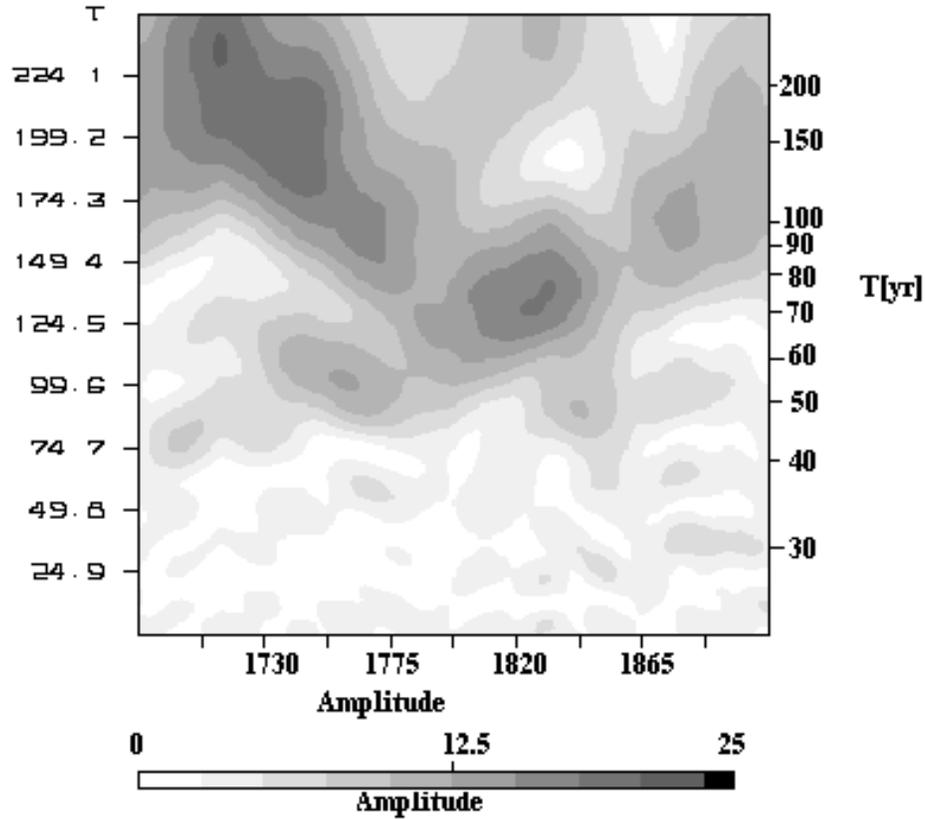

Fig. 05 The MWTRPP amplitude map of the Group sunspot number (*GSN*) series (AD 1610-1995)

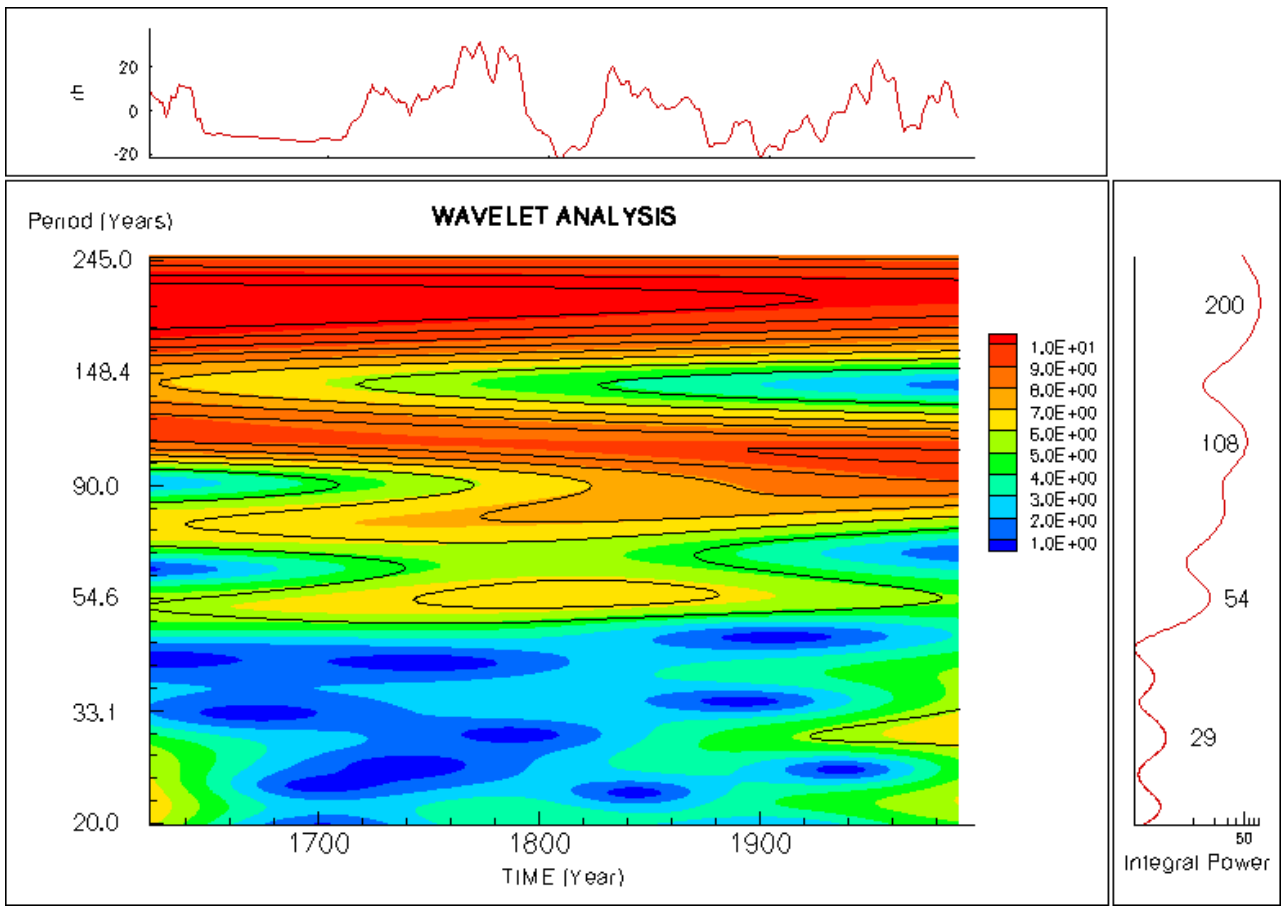

Fig. 06 WA amplitudes of the Group sunspot number (*GSN*) series (AD 1610- 1995)

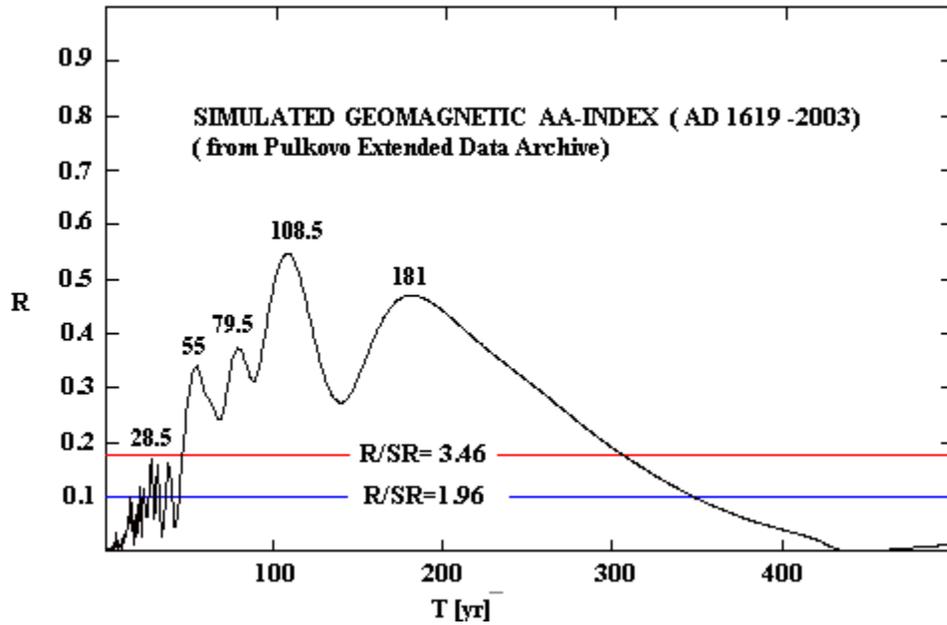

**Fig. 7** T-R corellogram of the simulated *aa*-index (*AA*) data series (AD 1619-2003, ESAI):

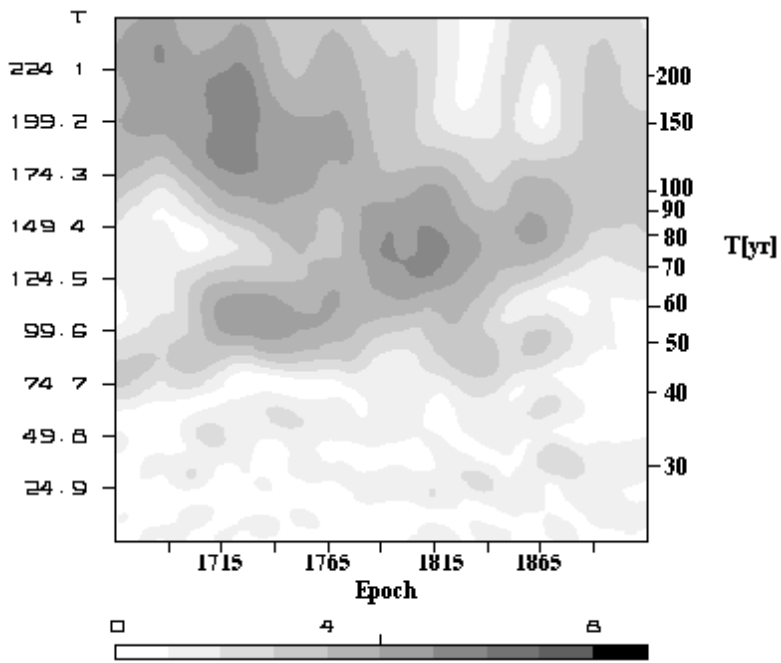

**Fig. 8** The MWTRPP amplitude map of the simulated geomagnetic *aa*-index (*AA*) data series (AD 1619-2003, ESAI)

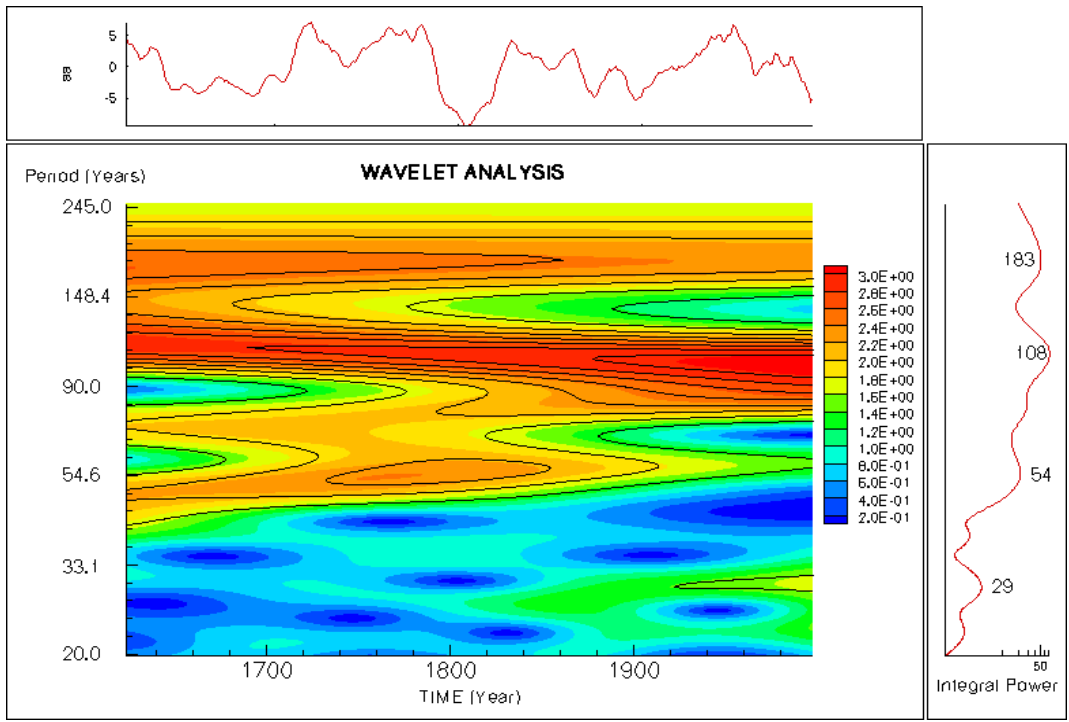

**Fig. 9** WA amplitudes of the simulated geomagnetic *aa*-index (*AA*) data series (AD 1619-2003, ESAI)

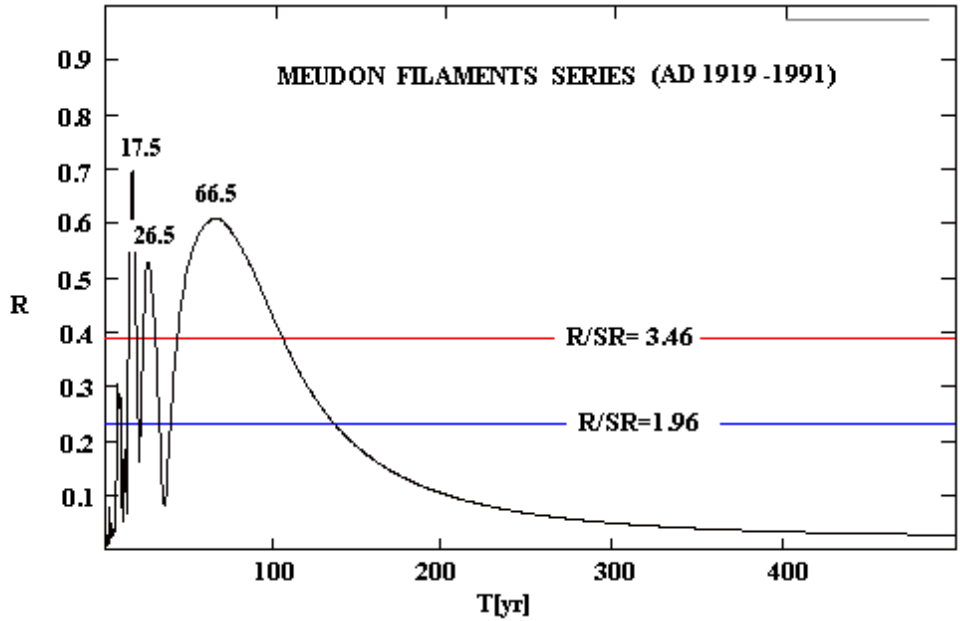

**Fig 10.** The T-R corellogram of the solar filaments number series from the Meudon catalogues (1919-1991)

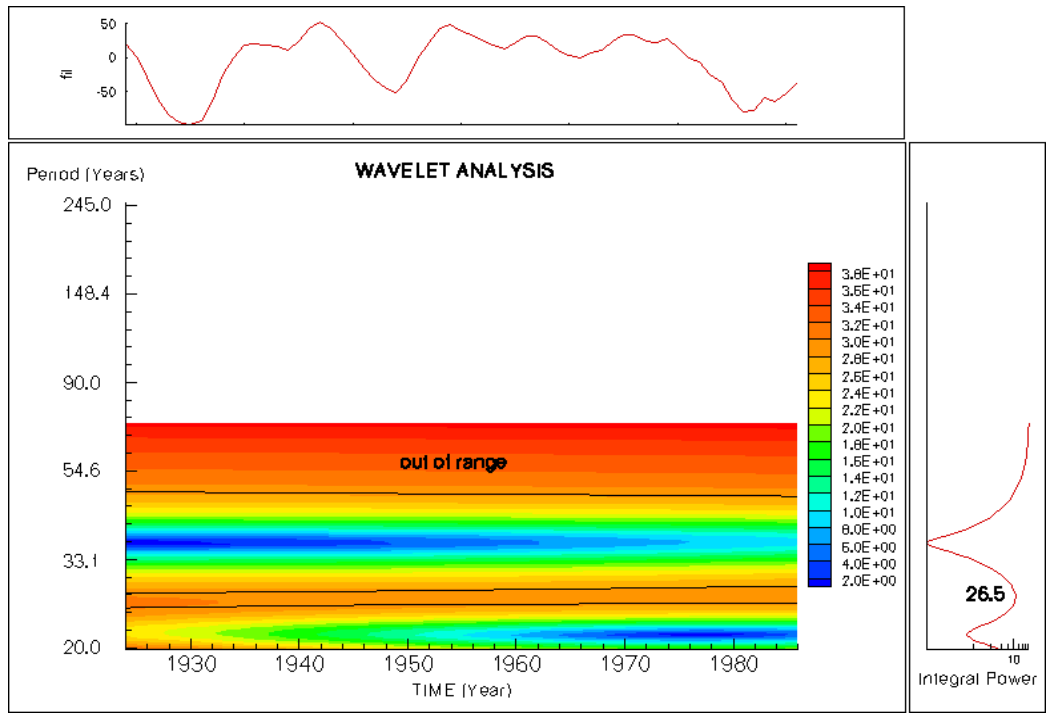

**Fig. 11. Wavelet analysis of the solar filaments number series from the Meudon catalogues (1919-1991)**

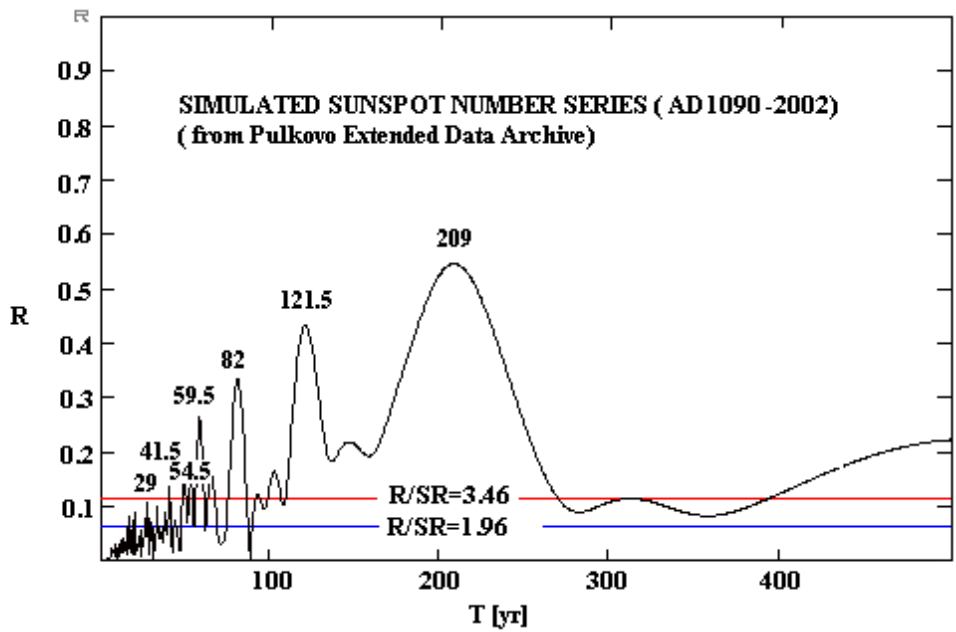

**Fig. 12  T-R corellogram of the whole simulated sunspot *Rsi* number series (AD 1090-2002, ESAI)**

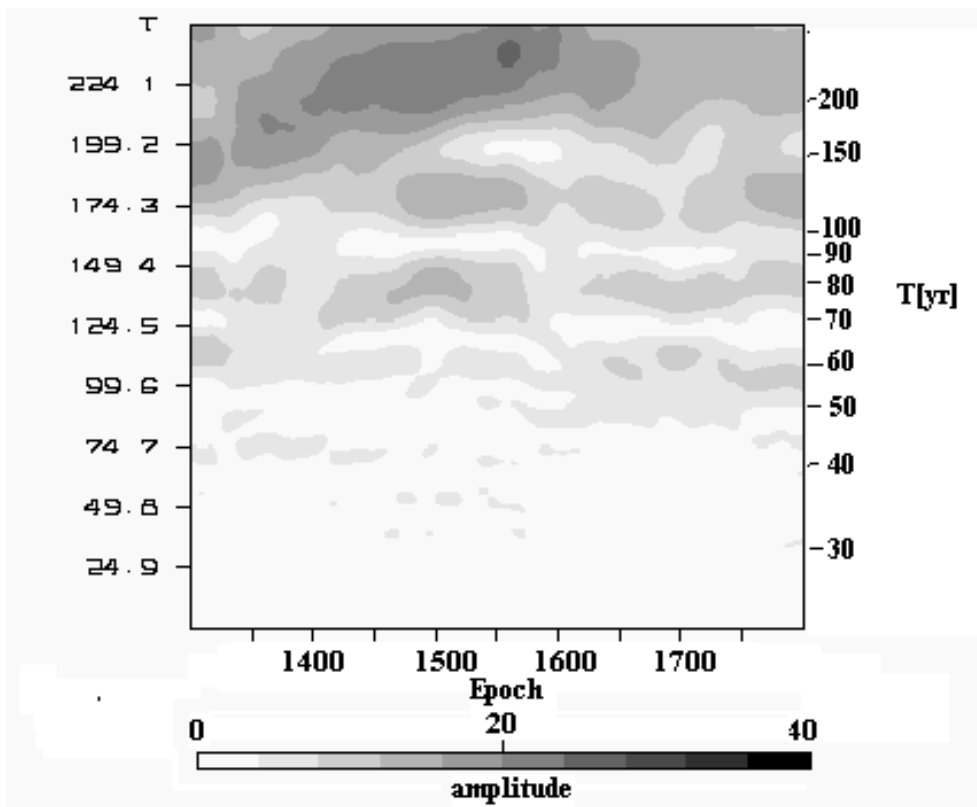

**Fig. 13** The MWTRPP amplitude map of the whole simulated sunspot *Rsi* number series (AD 1090-2002, ESAI)

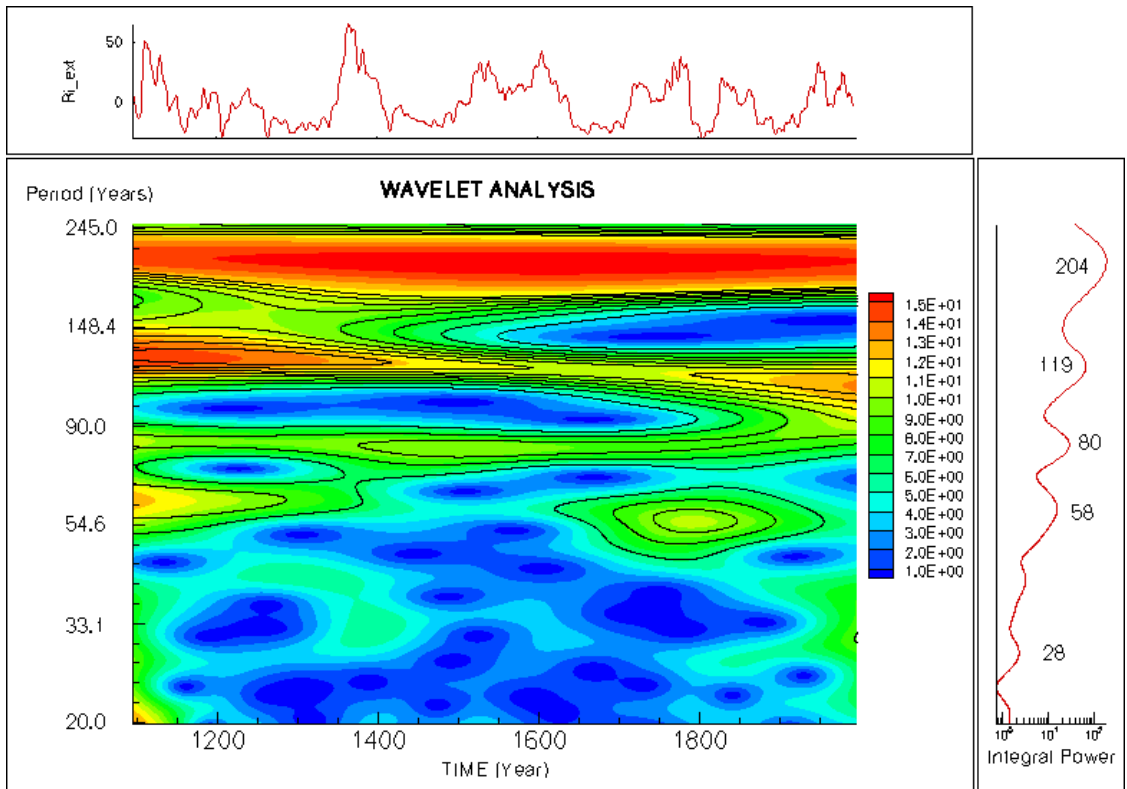

**Fig. 14** WA amplitudes of the whole simulated sunspot *Rsi* number series (AD 1090-2002, ESAI)

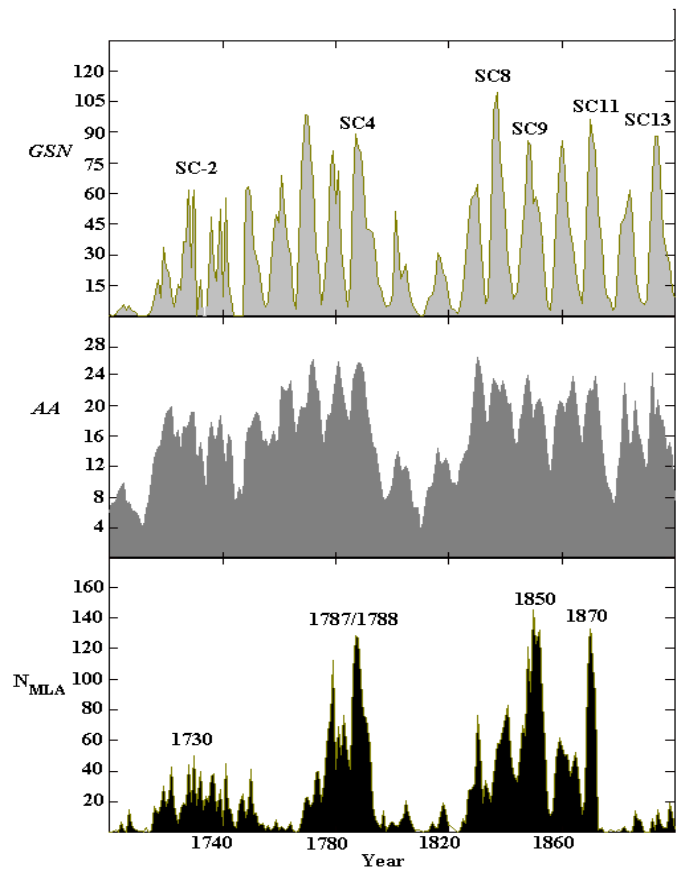

**Fig. 15 The sunspot (*GSN*), geomagnetic (*AA*) and MLA activity($N_{MLA}$) (AD 1700-1900)**